# DØ Data Handling Operational Experience

K. Yip
*Brookhaven National Laboratory, Upton, NY 11973*
A. Baranovski, D. Bonham,L. Carpenter, L. Lueking, W. Merritt, C. Moore, I. Terekhov, J. Trumbo, S. Veseli, J. Weigand, S. White
*FNAL, Batavia, IL 60510*
R. Brock
*Michigan State University, East Lansing, MI 48823*

We report on the production experience of the DØ experiment at the Fermilab Tevatron, using the SAM data handling system with a variety of computing hardware configurations, batch systems, and mass storage strategies. We have stored more than 300 TB of data in the Fermilab Enstore mass storage system. We deliver data through this system at an average rate of more than 2 TB/day to analysis programs, with a substantial multiplication factor in the consumed data through intelligent cache management. We handle more than 1.7 Million files in this system and provide data delivery to user jobs at Fermilab on four types of systems: a reconstruction farm, a large SMP system, a Linux batch cluster, and a Linux desktop cluster. In addition, we import simulation data generated at 6 sites worldwide, and deliver data to jobs at many more sites. We describe the scope of the data handling deployment worldwide, the operational experience with this system, and the feedback of that experience.

## 1. INTRODUCTION

The DØ Collaboration includes over 600 physicists from 80 institutions in 18 countries and a reliable, sophisticated data handling system is an important element to the experiment. The current running period, nominally referred to as Run 2a, started in 2002 is anticipated to extend through mid 2004. The Run 2a detector has approximately 1 Million channels and a Raw event size of 250 kByte with an online data-taking rate averaging 25 Hz. The estimated data totals for this phase of Run 2 are 1.2 x $10^9$ recorded triggers which amounts to 1.2 PBytes, including reconstructed and secondary data sets. In addition to this data, D0 has six Monte Carlo processing centers worldwide which will generate an additional 300 TB of information. The Run 2b phase of the experiment, starting in 2005, is anticipated to generate a total of over 1 PByte per year and will continue until the LHC begins to achieve results.

The DØ data handling operation is successful due to the scalable design, a dedicated staff at FNAL, and contributions made by the worldwide collaboration. The network-based hardware architecture has enabled the system to grow beyond the Fermilab boundaries, and the SAM Data Handling system software [1,2] has enabled a distributed computing system providing transparent access to the data at dozens of sites. The Enstore Mass Storage System [3,4] has provided robust and reliable storage and access of data to the tape archive. Monte Carlo production has been on-going for the last 4 years almost exclusively using non-FNAL resources and this has enabled the computing and personnel resources at Fermilab to be focused on processing the data originating from the d0 detector.

## 2. COMPUTING ARCHITECTURE

### 2.1. Computing Hardware

The computing hardware is connected through a high speed network infrastructure. The summary of the components of the hardware is given in Figure 1. At Fermilab, the high speed network provides more than 150 MBps of transfer capacity among the many components of the system. The major FNAL components consist of the following:

- Online data acquisition
- Robotic tape storage libraries
- Reconstruction farm
- Central analysis system
- Central Analysis Backend (CAB) compute server farm, and
- CLueD0 desktop cluster.

Although these systems represent the majority of the experiment's computing resources, significant facilities are being employed at sites beyond Fermilab. Event simulation and processing is performed almost exclusively at remote centers including the IN2P3 Computing Center, Lyon (France), Imperial College, London (UK), Lancaster University (UK), Manchester University (UK), Prague (Czech Republic), NIKHEF in Amsterdam (Netherlands), and the University of Texas Arlington, Arlington TX (USA). Additional future sites for data reconstruction and analysis are being commissioned at GridKa, Forschungszentrum Karlsruhe (FZK Germany), the University of Michigan, and several other locations.

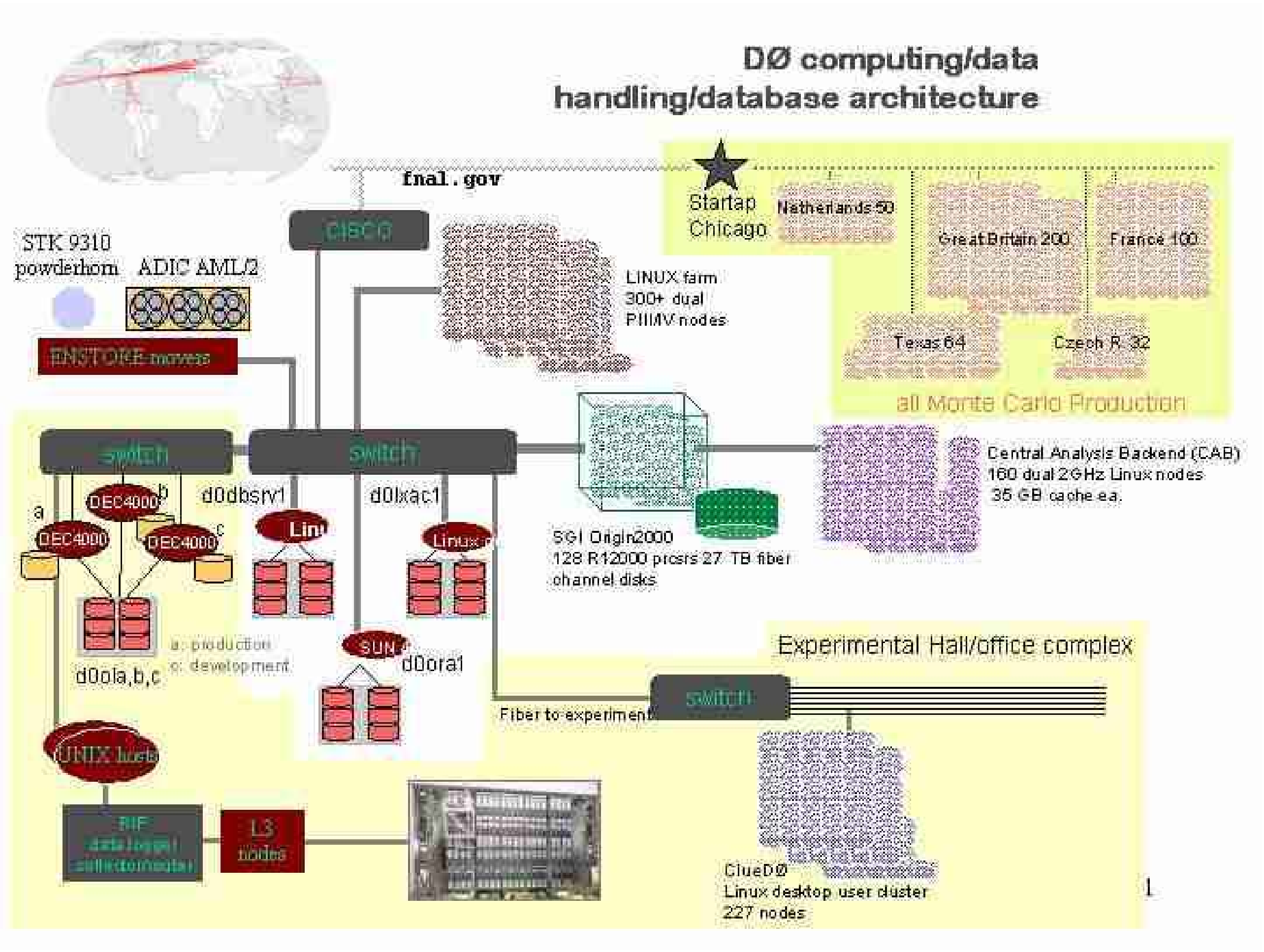

Figure 1: The DØ computing hardware layout at Fermilab showing the network connections for the systems described in the text.

## 2.2. The SAM Data Handling System

The SAM Data Handling system provides a sophisticated set of tools for storing, cataloguing and delivering data throughout the DØ collaboration. The system has a flexible and scalable distributed model with the architecture based on CORBA. It has been field hardened with over three years of production experience, and the system is reliable and fault tolerant. Virtually any local batch system (LBS) can be used with SAM and it has been functioning with many LBS's including LSF[5], PBS [6], Condor [7], FBS[8], and even Sun ONE Grid Engine (SGE)[9]. Adapters for many storage systems are also available including Enstore, HPSS [10], and TSM[11]. The system utilizes many transfer protocols including cp, rcp, scp, encp, bbftp, and GridFTP.

The system provides data management at the site and wide area level, but also provides extensive cache management for local compute resources. The collection of services which are required to provide the needed functionality for compute systems is referred to as a *SAM Station*. These servers manage data delivery and caching for a node or cluster. The node or cluster hardware which is managed by these services is also referred to as a SAM Station.

SAM is useful in many cluster computing environments including SMP and distributed clusters. SAM stations have been deployed on dedicated compute servers, as well as on a desktop cluster. The system has options for worker nodes on a private network (PN), and can manage distributed cache or a common cache on NFS shared disks.

## 3. OPERATIONAL STATISTICS

### 3.1. Overview

Table 1 provides a summary of the DØ SAM data management operation. Stations have been deployed across the collaboration, as shown in Figure 2. Over the last year over 4 Million files have been delivered by the system representing over 1.2 PByte of information. The integrated plots for these two metrics are shown in Figure 3. A few of the larger, and most active, station installations are overviewed in Table 2. Stations have broadly divergent configurations and usage patterns, depending on site practices and policy. They are tailored for particular activities and optimized by setting various station, group, and caching policy parameters.

Table 1: DØ Data Handling Statistics

| Statistic | Number/Size |
|---|---|
| Registered Users | 600 |
| Number of SAM Stations | 56 |
| Registered Nodes | 900 |
| Total Disk Cache | 40 TB |
| Number of Files - Physical | 1.2 M |
| Number of Files - Virtual | 0.5 M |
| Robotic Tape Storage | 305 TB |

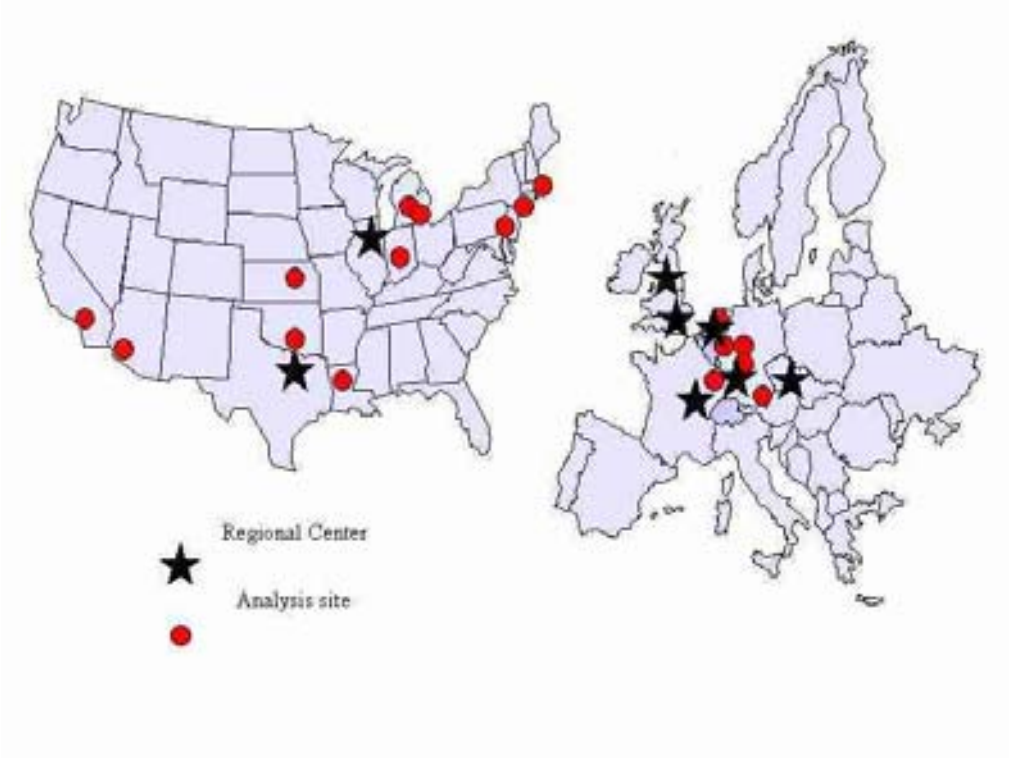

Figure 2: SAM station deployment with stars representing Regional or Monte Carlo production centers. Dots represent analysis stations.

An important component of the system has been the robotic tape storage facility at Fermilab managed with Enstore. Figure 2 charts the daily data into (+) and out of (-) this facility with typical storage of 1 TB per day, and reading 2 TB or more per day

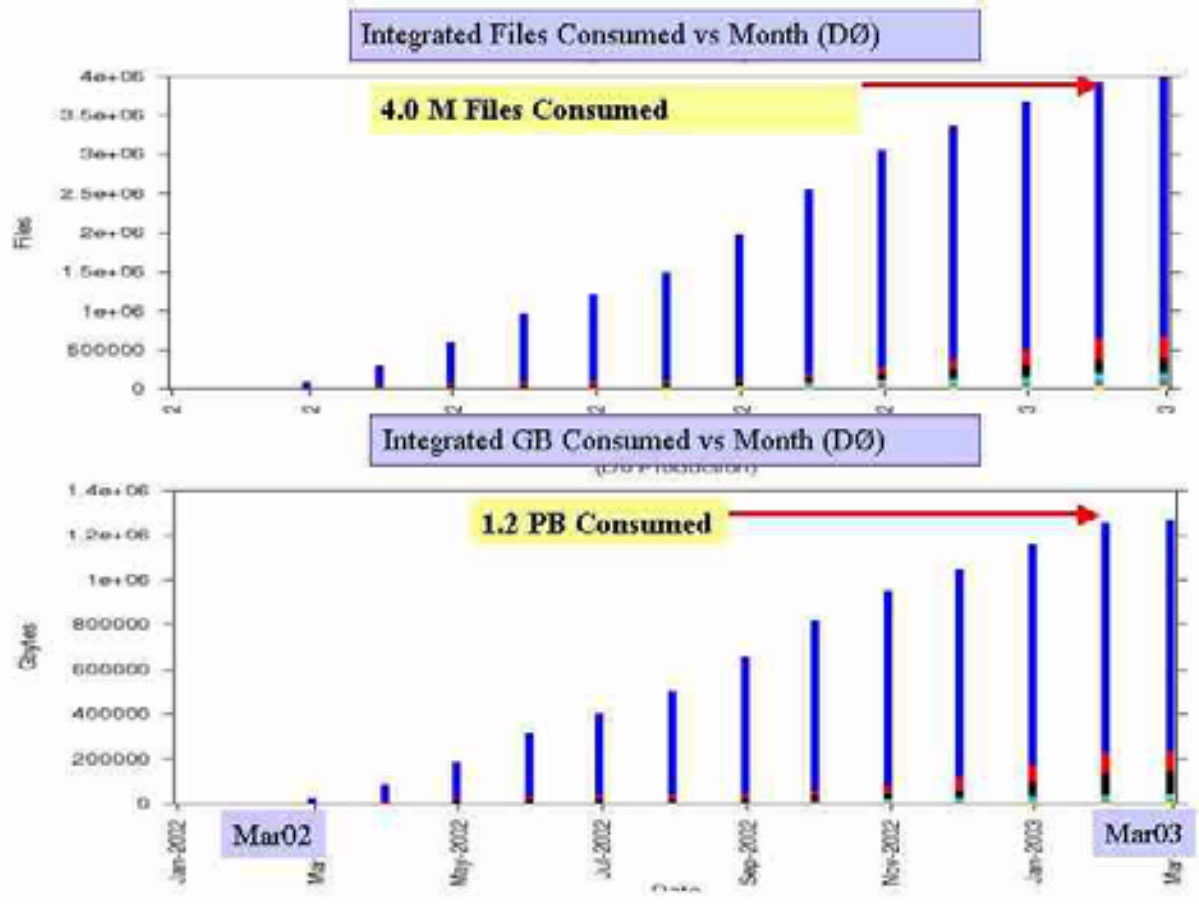

Figure 3: Data consumed in terms of files, and GBytes, over the period March 2002 to March 2003.

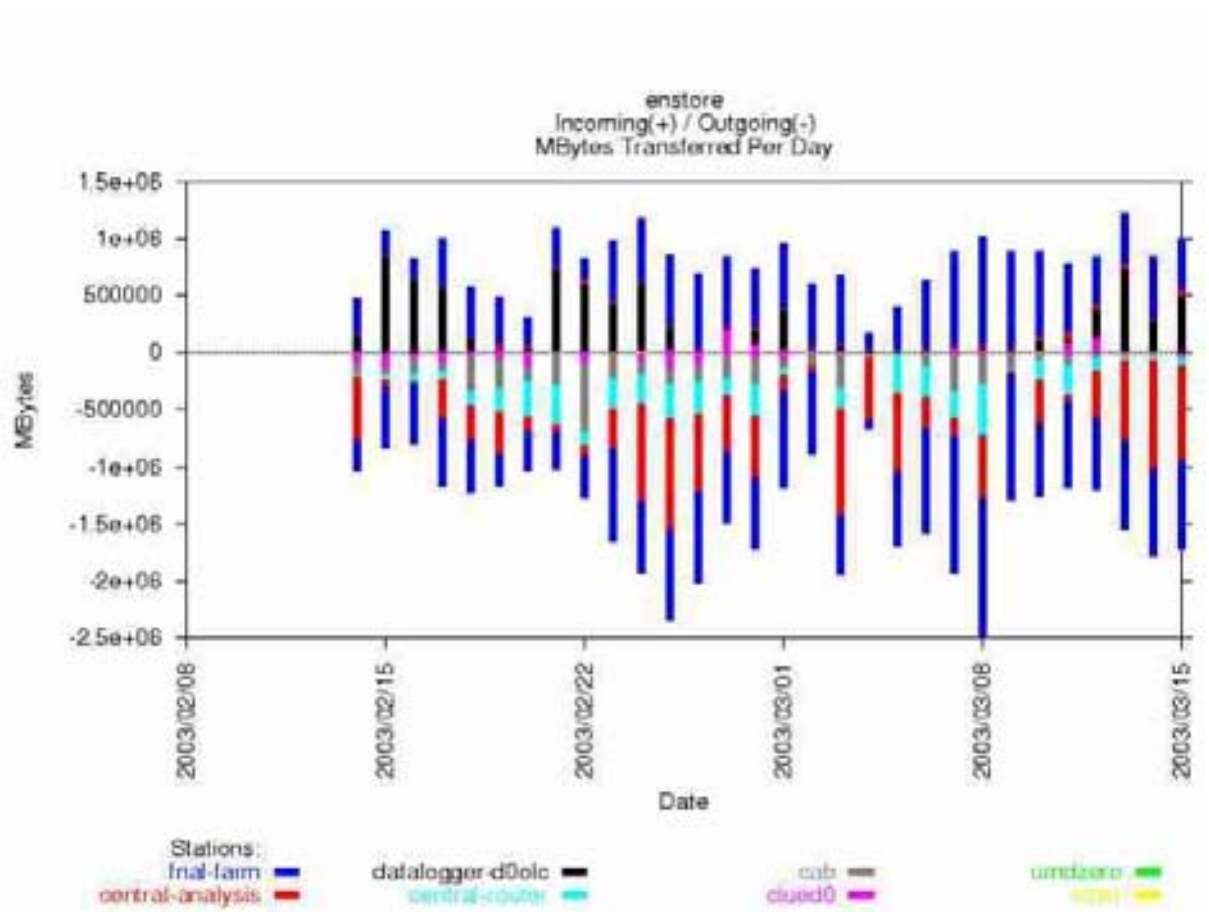

Figure 4: Data written (+) and read(-) to and from the Enstore system per day.

## 3.2 Specific Station Statistics

The operation of each station can be characterized by several metrics including the incoming flow of data, the outgoing flow of data, the consumption of data local to the station, and the CPU utilization. Figure 5 and 6 include charts for daily data movement for four of the systems in table 2: Central-Analysis, FNAL-Farm, CLueD0, and CAB. In figure 5, the data actually consumed by processes is shown with the peak days indicated. Figure 6 shows, for the same systems, over the same time period, the data delivered to the station (+) or data sent from the station (-).

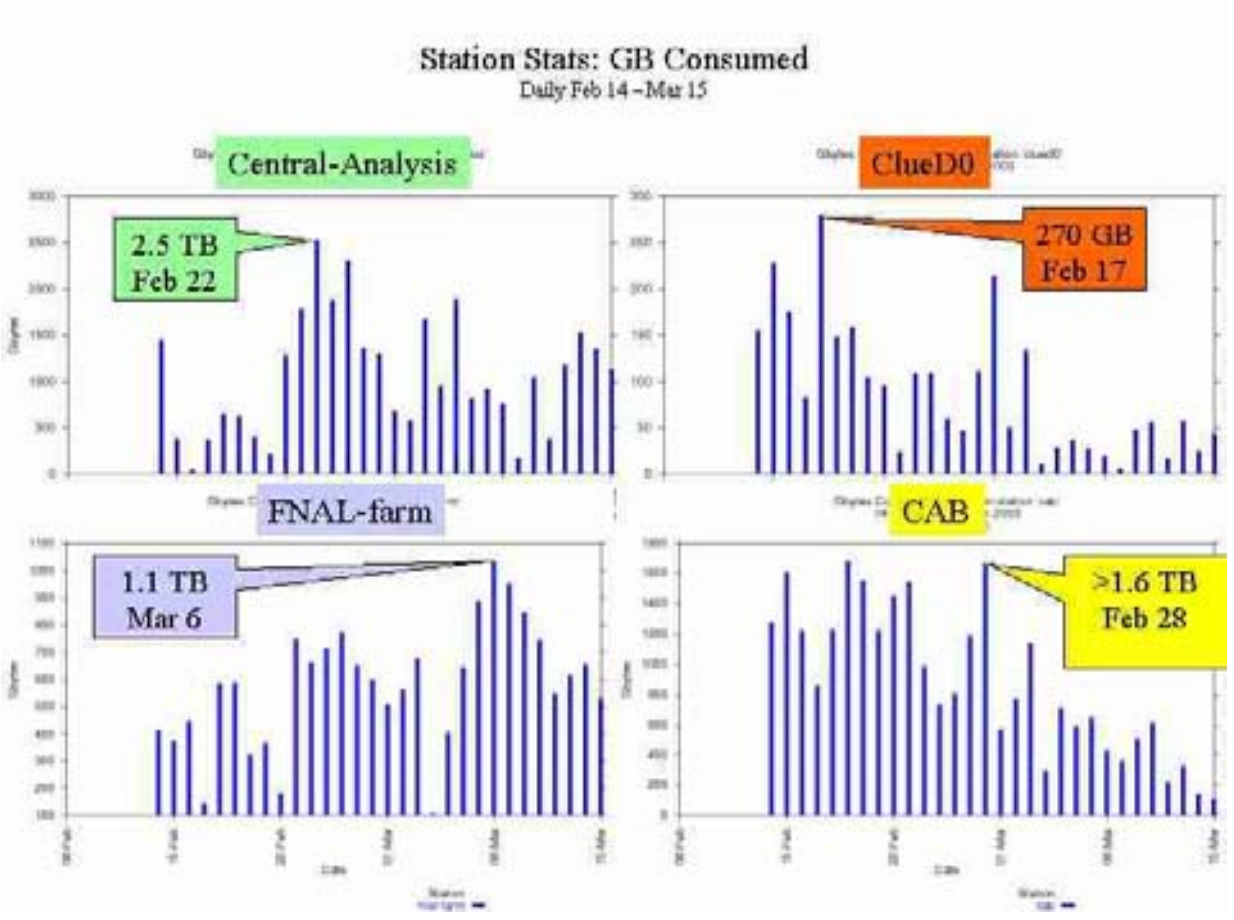

Figure 5: Data consumed on Central Analysis, ClueD0, Reconstruction Farm, and CAB systems.

By comparing the numbers for the two sets of figures one can determine the effectiveness of the cache management for various modes of operation. For example, on the Central-analysis station we observe that on a peak day, 2.5 TB were consumed while only 1 TB were brought into the system. The pattern of usage on Central Analysis is mostly analysis type activities, and the caching is extremely useful.

Table 2: Overview of SAM Stations in the DØ data handling system.

| Name | Location | Nodes/CPU | Cache | Use/Comments |
|---|---|---|---|---|
| Central-analysis | FNAL | 128 SMP*, SGI Origin 2000 | 14 TB | Analysis & D0 code development |
| CAB (CA Backend) | FNAL | 160 dual 1.8 GHz | 6.2 TB | Analysis and general purpose |
| FNAL-Farm | FNAL | 100 dual 0.5-1.0 GHz +240 dual 1.8 GHz | 3.2 TB | Reconstruction |
| ClueD0 | FNAL | 50 mixed PIII, AMD. (may grow >200) | 2 TB | User desktop, General analysis |
| D0 Karlseruhe | GridKa Karlsruhe, Germany | 1 dual 1.3 GHz gateway, >160 dual PIII & Xeon | 3 TB NFS shared | General/Workers on PN. Shared facility |
| D0 Umich (NPACI) | U..Michigan Ann Arbor, MI | 1 dual 1.8 GHz gateway, 100 x dual AMD XP 1800 | 1 TB NFS shared | Re-reconstruction. workers on PN. Shared facility |
| Robotic Tape Storage | Worldwide | Mostly dual PIII, Xeon, and AMD XP | Varied | MC production, gen. analysis |

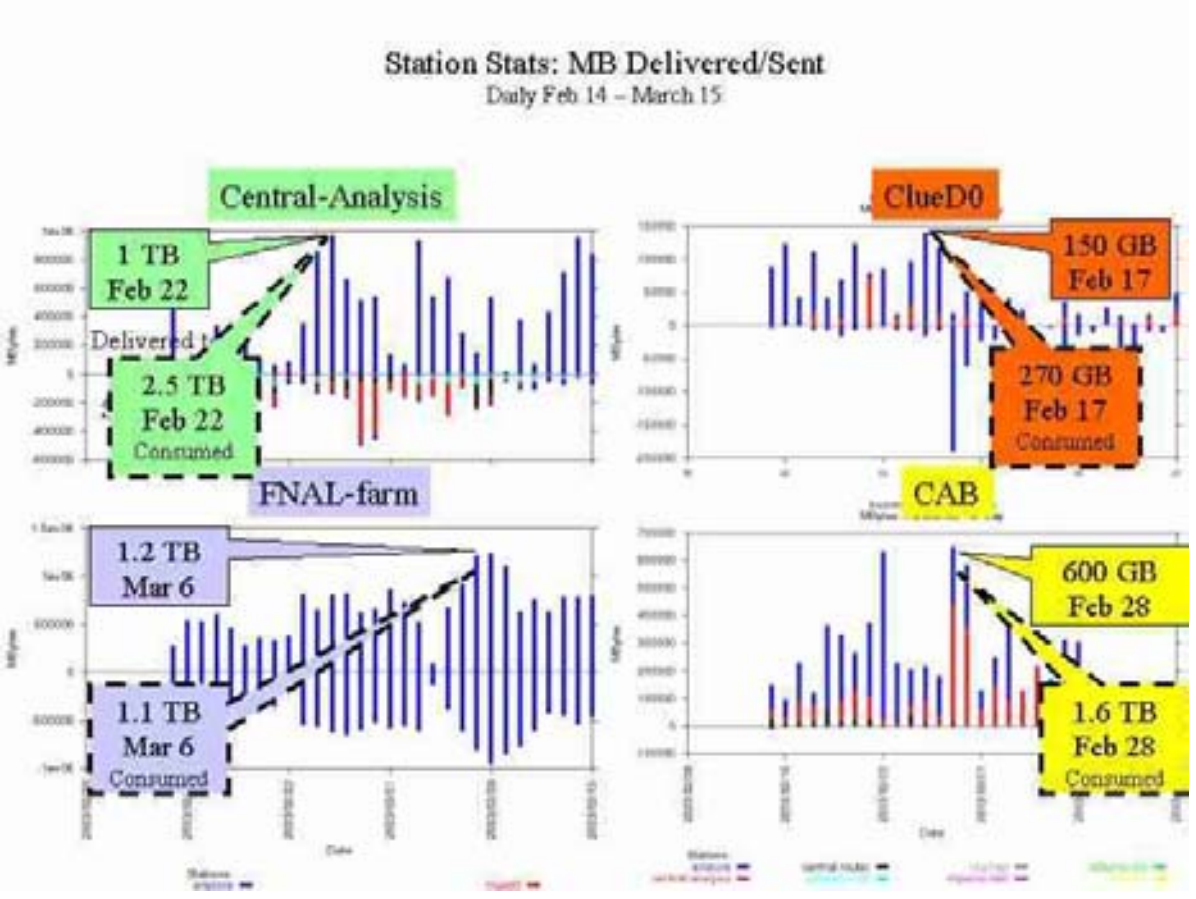

Figure 6: Data movement in to and out of Central Analysis, ClueD0, FNAL-Farm, and CAB SAM stations.

On the other hand, if we look at the same numbers for the FNAL-Farm, we observe that on a peak day 1.2 TB were delivered to the system, while only 1.1 TB were actually consumed. This is quite typical for the reconstruction mode of operation and the cache management plays a different role than in the analysis activity.

### 3.3 Data Handling Beyond Fermilab

Beyond Fermilab, SAM stations are an effective way for the collaboration to transparently access data. SAM stations are configurable to statically route data throughout the system. Figure 7 illustrates how this might be done to move data from/to remote sites, through a particular remote station, to/from stations at Fermilab. Such routing is established so that regional centers throughout the US and Europe can be employed as caching centers, and network hubs for institutions within the region.

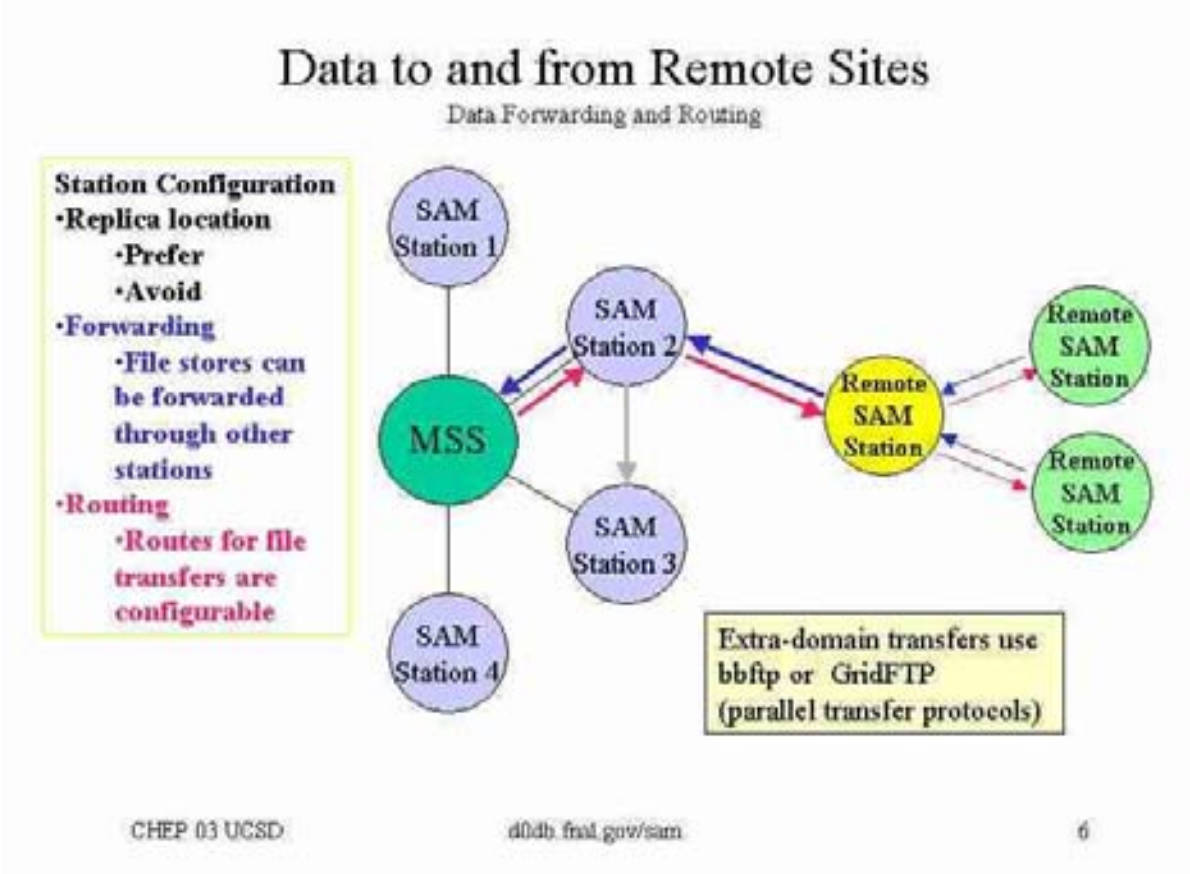

Figure 7: SAM stations can be configured with static routes to enable forwarding files from sites to particular storage systems, or replicating files routed through designated stations.

Several such centers have been identified and GridKa[12] is the prototype facility to test many of these ideas. Throughout the past 9 months data has been systematically pulled to the SAM station in Karlsruhe and cached. This was processed and certified to produce identical results with data sets maintained at FNAL. This facility has successfully processed data used to generate results for the winter (March 2003) physics conferences. We consider this prototype to be a huge success and continue to move ahead with the development of the regional center at this site, and others. Additional information with regard to the DØ Regional Analysis Center concepts can be found at [13].

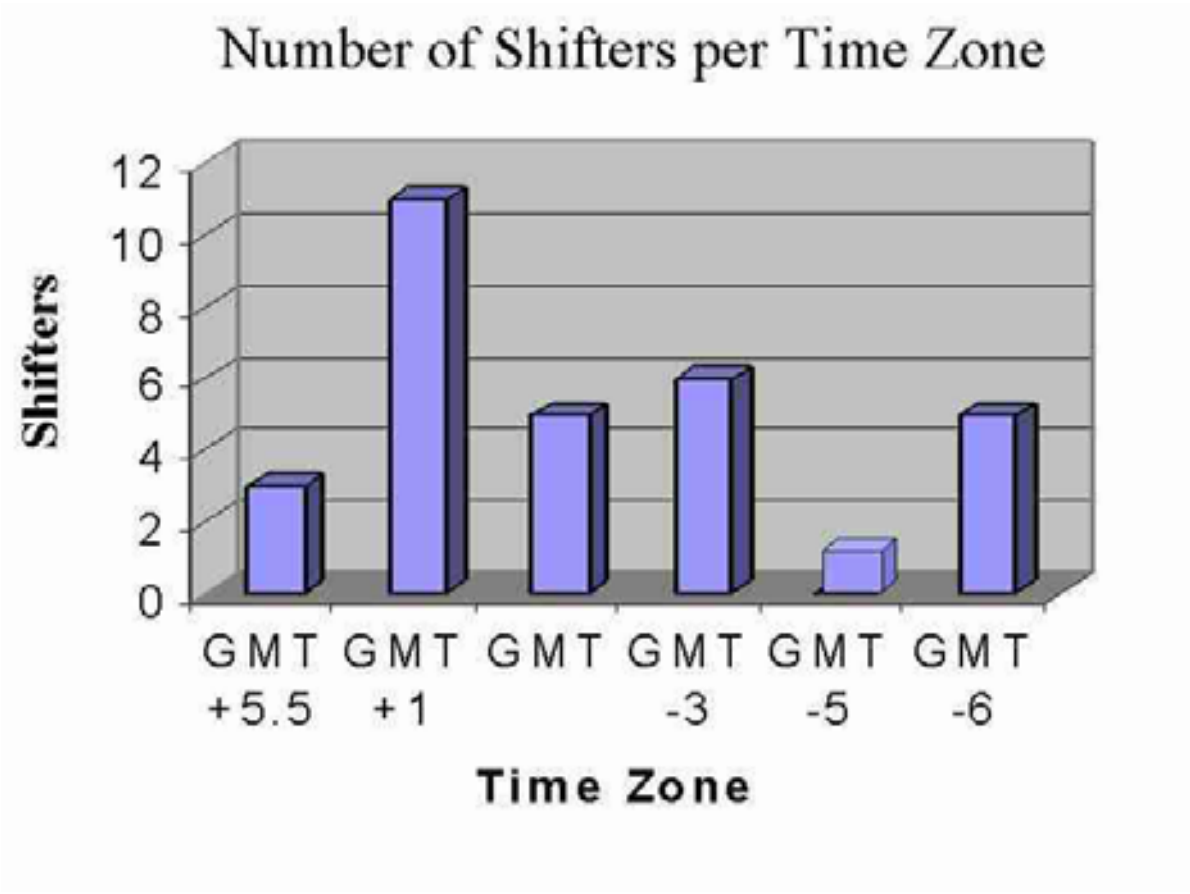

Figure 8: Number of shift personnel per time zone.

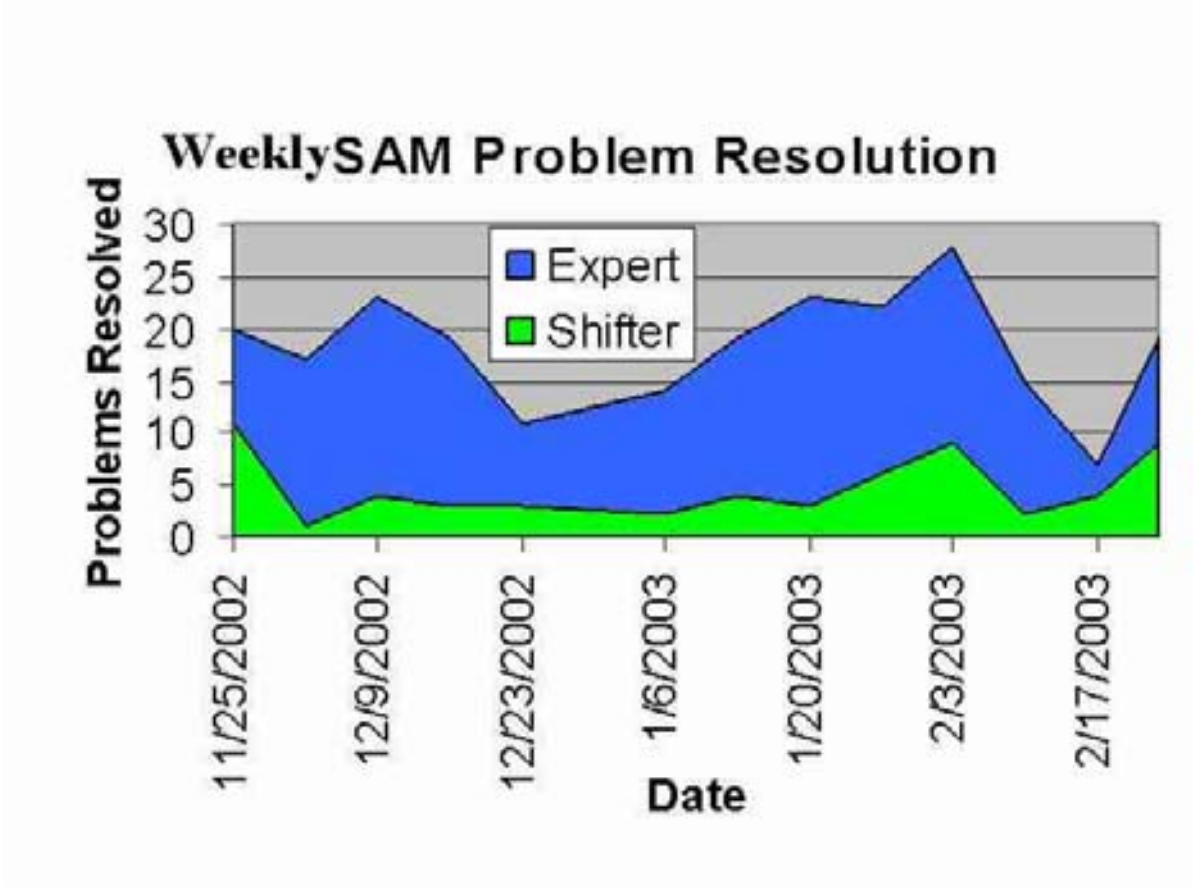

Figure 9: Number of problem resolutions per week. The average is about 3 per day with greatly varying in complexity.

## 4. CHALLENGES AND FUTURE PLANS

### 4.1 . Major Challenges

Getting SAM to meet the needs of DØ in the many configurations is, and has been, an enormous challenge. File corruption issues were creating problems for analysis and a method of calculating CRC values and checking the file each time it is moved have resolved this. Preemptive distributed caching is prone to race conditions and log jams; these issues have been solved by SAM station improvements. Private networks sometimes require “border” naming services to enable these configurations to be usable. An NFS shared cache configuration is now available to provide additional simplicity and generality, although at the price of scalability due to the star configuration.

Global routing is now enabled and this enables static routes to be configured (as described earlier). Installation procedures for the station servers have been quite complex, but they are improving and we plan to soon have “push button” and even “opportunistic deployment” installs. Lots of details with opening ports on firewalls, OS configurations, registration of new hardware, and so on, have been resolved. Username clashing issues occur frequently because of the many systems and sites involved, so we are moving to GSI[14] and the use of Grid Certificates for authentication. Interoperability with many MSS’s have been achieved. Network attached files are now enabled so the file does not need to move to the user, but can remain on a remote server and accessed through a network protocol, such as RFIO.

### 4.2. SAM Shift Operation

One of the important, but often overlooked, aspects of a system of this size and complexity is the day-to-day operation. D0 has had in place since the summer of 2001, a program of offline shift personnel composed of trained collaborators from around the globe. There are 27 shifters in 6 time zones with the distribution shown in Figure 8. With some additional online expert contributions, the system is covered on a 24/7 basis. Training this group is quite difficult since many are rarely at FNAL, so extensive web-based documentation and monitoring is essential. In addition to the experiment shifters, SAM developers act as experts to field problems which the front-line shifters cannot. The charts in Figure 9 show that the problem rate is typically about 20 per week, and the percentage of problems answered by general shifters continues to improve over time.

### 4.3. The SAM Grid Project

The SAM Data Handling system is envisioned to be just one component in a larger computing Grid system being built for the Run 2 experiments. This broader system, called SAM-Grid, includes Job and Information Management (JIM)[15] components which are being built within a project sponsored by PPDG [16]. This system uses standard grid middleware to build a fully functional grid, and is being deployed throughout the DØ and CDF collaborations over the summer of 2003. This is an exciting enhancement to the services already provided by the existing SAM system and we eagerly anticipate its full functionality.

## 5. SUMMARY

The DØ Data Handling operation is a complex system involving a worldwide network of infrastructure and support. SAM provides flexible data management solutions for many hardware configurations, including clusters in private networks, shared NFS cache, and distributed cache. It also provides configurable data routing throughout the install base. The software is stable and provides reliable data delivery and management to production systems at FNAL and beyond. Many challenging problems have been overcome to achieve this goal.

Support is provided through a small group of experts at FNAL, and a network of shifters throughout the world. Many tools are provided to monitor the system, detect and diagnose problems. The system is continually being improved, and additional features are planed as the system moves beyond data handling to complete Grid functionality in the SAM-Grid project.

## Acknowledgments

We would like to thank the Fermilab Computing Division for its ongoing support of SAM, especially the CCF, CEPA, and D0CA Departments. We would like to thank everyone at DØ who has contributed to this project, and the many important discussions we have had there. This work is sponsored by DOE contract No. DE-AC02-76CH03000. Additional funding for Grid related SAM work is provided by the DOE SciDAC program through Dzero's association with the Particle Physics Data Grid (PPDG) collaboration.

Major contributions to the system and the operational coordination have come from Amber Boehnlein, the DØ Offline Computing Leader, Iain Bertram (Lancaster University, UK), Jianming Qian (UM), Rod Walker (IC), and Vicky White (FNAL). Several FNAL Computing Division Departments have been especially instrumental in the deployment and operation of the system. These include the Computing and Communication Fabric Dept. (CCF), in particular the Enstore team, the Core Support Services (CSS), in particular the Database Support Group (DSG), and Farms Support Group, and the Computing and Engineering for Physics Applications (CEPA) Database Applications Group (DBS) for database support, and the DØ Department and DØ Operations team at Fermilab.

Also, the CAB and CLueD0 administrators and support teams, Sam Station Administrators, and SAM Shifters Worldwide are among the many individuals in the experiment who make the system work. Monte Carlo production teams are located in Lancaster, UK, Imperial College London, UK, Prague in the Czech Republic, The University of Texas Arlington, the IN2P3 Computing Center in Lyon, France, and at NIKHEF, Amsterdam, in the Netherlands. We thank the staff at the GridKa Regional Analysis Center at Karlsruhe, Germany, and in particular Daniel Wicke (Wuppertal), Christian Schmitt (Wuppertal), and Christian Zeitnitz (Mainz) for their continued support for the system.

## References

[1] The SAM team, A. Baranovski, L. Loebel Carpenter, , L. Lueking , W. Merritt, C. Moore, I. Terekhov, J. Trumbo, S. Veseli, S. White, http://d0db.fnal.gov/sam

[2] Baranovski, et. al., "SAM Managed Cache and Processing for Clusters in a Worldwide Grid-Enabled System", FERMILAB-TN-2175, May 2002. (http://d0db.fnal.gov/sam_talks/talks/20021021-cluster/ClusterComputing2002.pdf)

[3] The Enstore home page, http://www-isd.fnal.gov/enstore

[4] J. Bakken, et. al. , This conference.

[5] The LSF Batch system, Platform Computing, http://www.platform.com/

[6] The Portable Batch System, http://http://www.openpbs.org

[7] The Condor project home page, http://www.cs.wisc.edu/condor/

[8] FBS is the Fermilab Batch System, http://www-isd.fnal.gov/fbsng

[9] Sun Grid Engine http://wwws.sun.com/software/gridware/

[10] High Performance Storage System, http://www.sdsc.edu/hpss/

[11] The Tivoli Storage Manager, http://www-3.ibm.com/software/tivoli/

[12] The GridKa project, http://hikwww2.fzk.de/grid

[13] L. Lueking, et. al., "D0 Regional Analysis Concepts," CHEP03, UCSD, La Jolla CA, March 24-28, 2003 , TUAT002.

[14] Grid Security Infristrustructure from the Globus project, http://www.globus.org/security/

[15] I. Terekhov, et. al., "Grid Job and Information Management for the FNAL Run II Experiments,". CHEP03, UCSD, La Jolla CA, March 24-28, 2003, TUAT001.

[16] The Particle Physics Data Grid co-laboratory project, http://www.ppdg.net